\documentclass[12pt]{article}
\usepackage{amsmath,amssymb}
\def\today{\ifcase\month\or
  January\or February\or March\or April\or May\or June\or
  July\or August\or September\or October\or November\or December\fi
        \space \number\year}
\def\versionno{M.~Postma}
\catcode`\@=11
{\count255=\time\divide\count255 by 60
\xdef\hourmin{\number\count255}
\multiply\count255 by-60\advance\count255 by\time
\xdef\hourmin{\hourmin:\ifnum\count255<10 0\fi\the\count255}}
\def\ps@draft{\let\@mkboth\@gobbletwo
    \def\@oddhead{}
    \def\@oddfoot{\hbox to 7 cm{\tiny \versionno
       \hfil}\hskip -7cm\hfil\rm\thepage \hfil {\tiny\draftdate}}
    \def\@evenhead{}\let\@evenfoot\@oddfoot}
\def\draftdate{\number\day.\number\month.\number\year\ \ \ \hourmin}

\catcode`\@=12
\newcommand{\thisfile}[1]{\renewcommand{\versionno}[0]%
{M.~Postma --- File: #1.tex}}
\input epsf
\def\be{\begin{equation}}
\def\ee{\end{equation}}
\def\bea{\begin{eqnarray}}
\def\eea{\end{eqnarray}}

\renewcommand\({\left(}
\renewcommand\){\right)}
\renewcommand\[{\left[}
\renewcommand\]{\right]}

\newcommand{\dd}{{\rm d}}

\newcommand{\odec}{\Omega_{\rm dec}}
\newcommand{\oosc}{\Omega_{\rm osc}}
\newcommand{\ophi}{\Omega_\phi}
\newcommand{\os}{\Omega_S}
\newcommand{\opt}{\Omega_{\rm pt}}
\newcommand{\ts}{\textstyle}

\newcommand\TeV{\,\mbox{TeV}}
\newcommand\GeV{\,\mbox{GeV}}

\newcommand\mpl{M_{\rm pl}}
\newcommand\Mpl{M_{\rm pl}}

\begin{document}

%\draft
\thisfile{low}

\vskip 1cm
\begin{center}
{\Large \bf Curvaton scenario with low scale inflation revisited
\vskip 0.2cm}
\vspace{0.4in}
{Marieke Postma}
\vspace{0.3in}

{\baselineskip=14pt \it 
The Abdus Salam ICTP, Strada Costiera 11, Trieste, Italy \\[1mm]}

\vspace{0.2in}
%$^b$ {\baselineskip=14pt \it address \\[1mm]}
%\vspace{0.2in}
\end{center}

\vspace{0.2cm}
\begin{center}
{\bf Abstract} 
\end{center}
\vspace{0.2cm} 
In its simplest form the curvaton paradigm requires the Hubble
parameter during inflation to be bigger than $10^8 \GeV$, but this
bound may be evaded in non-standard settings.  In the heavy curvaton
scenario the curvaton mass increases significantly after the end of
inflation.  We reanalyze the bound in this set up, taking into account
the upper bound on the curvaton mass from direct decay. We obtain $H_*
> 10^8 \GeV$ if the mass increase occurs at the end of inflation, and
$H_* > 10^{-14} \GeV$ if it occurs just before nucleosynthesis. We
then discuss two implementations of the heavy curvaton.  Parameters
are constrained in these explicit models, and as a result even
obtaining TeV scale inflation is hard to obtain.

\newpage

%%%%%%%%%%%%%%%%%%%%%%%%%%%%%%%%%%%%%%%%%%%%%%%%%%%%%%%%%%%%%%%%%%%%%%%%%%
%%%%%%%%%%%%%%%%%%%%%%%%%%%%%%  section 1 %%%%%%%%%%%%%%%%%%%%%%%%%%%%%%%%
%%%%%%%%%%%%%%%%%%%%%%%%%%%%%%%%%%%%%%%%%%%%%%%%%%%%%%%%%%%%%%%%%%%%%%%%%%

\section{Introduction}

In the curvaton scenario not the inflaton but some other light field
--- the curvaton field --- is responsible for the observed density
perturbations~\cite{curv}.  In the post inflationary epoch the
curvaton oscillates around the origin of its potential in a radiation
dominated background.  It is during this period that the curvaton
contribution to the total energy density grows and the isocurvature
perturbations of the curvaton field are converted into adiabatic ones.
The resulting perturbation spectrum is Gaussian if the curvaton
accounts for at least one percent of the total energy density at the
time of decay.

The great merit of the curvaton scenario is that it liberates models
of inflation~\cite{liberated}.  Both the scale of inflation, and the
flatness of the inflaton potential are decoupled from the observed
magnitude and scale dependence of the cosmic microwave background
(CMB) spectrum.  This means in particular that inflation can take
place at a much lower scale than the customary $H_* \sim
10^{10}-10^{14} \GeV$.

Many models of inflation have been constructed, in which the inflaton
potential is generated by the same supersymmetry (SUSY) breaking
mechanism that operates in the vacuum, and the scale of inflation is
naturally $H_* \sim 10^3 \GeV$~\cite{low_inflation,NOI,disformal}.
However, as was shown in Ref.~\cite{lyth} the curvaton scenario in its
simplest form cannot accommodate such low inflationary scales either.
The reason is that the lower the scale of inflation, the lower is the
initial curvaton energy density and the shorter the period of
oscillations.  Only for large enough Hubble constants during
inflation, $H_* > 10^7 \GeV$, can the curvaton come to dominate the
energy density (and decay) before the epoch of nucleosynthesis.

To rescue low scale models of inflation it is necessary to go beyond
the simplest curvaton scenario.~\footnote{See Ref.~\cite{postma_irs}
  for a discussion of low scale inflation in the context of the
  inhomogeneous reheating scenario.} One way to reconcile low scale
inflation with the curvaton scenario is to consider curvaton
oscillations in a background dominated by a stiff fluid instead of
radiation~\cite{tracking}. In Refs.~\cite{lyth,matsuda} the ``heavy
curvaton'' scenario was proposed, in which the curvaton mass increases
significantly after the end of inflation.  In this paper we reanalyze
the bound on the Hubble constant during inflation within this
scenario.  As the curvaton mass increases so does its decay rate.
This leads to an upper bound on the curvaton mass at the time of
curvaton decay; increasing the mass beyond this bound will lead to a
rapid decay of the curvaton condensate. This constraint was not taken
into account in previous works.  We obtain the bound on the
inflationary scale in the heavy curvaton scenario $H_* > 10^8 \GeV$ if
the mass increase occurs right after the end of inflation, and $H_* >
10^{-14} \GeV$ if it occurs just before big bang nucleosynthesis
(BBN).

In the next section we start with a review of the bound on the Hubble
scale during inflation in the simplest curvaton scenario.  How this
bound can be evaded is discussed in turn in section~\ref{s_evading}.
In particular, the heavy curvaton scenario is detailed.  The rest of
the paper is devoted to two specific implementations of the heavy
curvaton.  In section~\ref{s_higgs} we discuss a model in which the
curvaton mass increases due to a direct coupling to a ``Higgs'' field.
In section~\ref{s_axion} we analyze a scenario in which the curvaton
is an axion. Its mass receives contributions from two different
non-renormalizable operators; one operator dominates during inflation,
whereas the other dominates afterwards.  Although theoretically an
inflationary scale as low as $H_* \sim 10^{-14} \GeV$ is possible, we
find that in these explicit models even low scale inflation with $H_*
\sim \TeV$ is hard to obtain.  The reason is that once an explicit
model is specified, there are relations between the various
parameters, which can no longer be varied independently.

%%%%%%%%%%%%%%%%%%%%%%%%%%%%%%%%%%%%%%%%%%%%%%%%%%%%%%%%%%%%%%%%%%%%%%%%%%
%%%%%%%%%%%%%%%%%%%%%%%%%%%%%%  section 2 %%%%%%%%%%%%%%%%%%%%%%%%%%%%%%%%
%%%%%%%%%%%%%%%%%%%%%%%%%%%%%%%%%%%%%%%%%%%%%%%%%%%%%%%%%%%%%%%%%%%%%%%%%%

\section{The bound on the inflationary scale}
\label{s_ss}

We define the curvaton density parameter $\Omega_\sigma \equiv
\rho_\sigma /\rho_{\rm tot}$, which gives the contribution of the
curvaton to the total energy density.  The curvaton density parameter
changes with time; at the onset of curvaton oscillations
\be
\oosc
= \frac{m_\sigma^2 \sigma_*^2}{H_{\rm osc}^2 \mpl^2},
\ee
with $m_\sigma$ the curvaton mass.  The subscripts 'osc' and '*'
denote the corresponding quantity at the onset of curvaton
oscillations, and during inflation at the time observable scales leave
the horizon, respectively. We have dropped the subscript $\sigma$ from
the curvaton density parameter to avoid notational cluttering.  If the
curvaton and the background energy density red shift at different
rates, the curvaton contribution to the total energy density changes,
according to
\be
\frac{\dd \Omega_\sigma}{\Omega_\sigma} = -\alpha(H) \frac{\dd H}{ H},
\ee
with
\be
\alpha(H) = \frac{2 (\omega_{\rm tot} -\omega_\sigma)}
{1+ \omega_{\rm tot}}.
\label{alpha}
\ee
Here $\omega_{\rm tot} = (\Sigma_i \omega_i \rho_i)/\rho_{\rm tot}$
with $p_i = \omega_i \rho_i$ the equation of state for each fluid.
The quantity $\alpha(H)$ changes if either the equation of state
parameter of the background or that of the curvaton changes. In
particular, when the curvaton energy density comes to dominate the
total energy density, $\omega_{\rm tot} \to \omega_{\sigma}$, the back
ground and curvaton red shift at the same rate and $\alpha(H) \to 0$.
The ratio $\Omega_\sigma$ remains less than unity as it should be.

Consider the simple case that $\alpha$ remains constant after the
onset of curvaton oscillations until the curvaton energy density
approaches the total energy density.  The generalization to a changing
$\alpha$, which occurs for example if the inflaton field decays after
the onset of oscillations, is straightforward; we will comment on it
later. Denote by $\alpha$ the constant value of $\alpha(H)$ when
$\rho_\sigma \ll \rho_{\rm tot}$.  Since $\alpha(H)$ vanishes in the
limit that the curvaton comes to dominate the energy density we can
write down the inequality:~\footnote{If $\Gamma_\sigma$ is much
  smaller than $H_{\rm dom}$, the Hubble constant at which the
  curvaton comes to dominate the energy density, this equality is
  satisfied by many orders of magnitude.  As a consequence the Hubble
  bound in Eq.~(\ref{bound}) is off (to weak) by an amount
  $(\Gamma_\sigma/H_{\rm dom})^{\alpha/2}$.  We will come back to this
  issue when discussing the heavy curvaton.}
\be
\odec \lesssim
\frac{m_\sigma^2 \sigma_*^2}{H_{\rm osc}^2 \mpl^2} 
\( \frac{H_{\rm osc}}{\Gamma_\sigma} \)^\alpha,
\label{rdecay}
\ee
with $\Gamma_\sigma$ the curvaton decay rate, and $\odec$ the curvaton
density parameter at the time of curvaton decay.  The observed
spectrum of density perturbations is obtained for~\cite{wands,WMAP}
\be
\sigma_* \approx  \odec q A^{-1} H_*,
\label{sigmastar}
\ee
with $A = 1.5 \pi \times 10^{-4}$.  The damping factor $q$ is defined
through $({\delta \sigma}/{\sigma})_{\rm dec} = q ({\delta
  \sigma}/{\sigma} )_*$. For a quadratic or flat potential $q=1$, but
it is different for other monomials. Combining
Eqs.~(\ref{rdecay},~\ref{sigmastar}) leads to the bound on the Hubble
constant during inflation:
\be
H_* \gtrsim \frac{H_{\rm osc} \mpl A}{m_\sigma q \odec^{1/2}}
\( \frac{\Gamma_\sigma}{H_{\rm osc}} \)^{\alpha/2}.
\label{bound}
\ee
The minimum value for the decay rate is $\Gamma_\sigma >
\max[\Gamma_{\rm grav},\, \Gamma_{\rm bbn}]$, with $\Gamma_{\rm grav}
\sim m_\sigma^3 /\mpl^2 $ the decay rate for gravitational decay, and
$\Gamma_{\rm bbn} \sim 4 \times 10^{-25}\GeV$ corresponds to decay
just before BBN.

The bound on $H_*$ is minimized in the limit that the curvaton decays
just when it comes to dominate the energy density.  The period of
curvaton oscillations has to be sufficiently long for domination to be
possible.  Since for a lower scale of inflation the initial curvaton
vacuum expectation value (VEV) is smaller, see Eq.~(\ref{sigmastar}),
this period becomes increasingly long in this limit.  It might seem
that the period of curvaton oscillations can be shortened, and
therefore the bound be lowered, if the curvaton decays when it is
still sub-dominant in energy and $\odec < 1$.  However, this advantage
is more than off set by the fact that to obtain the right size of
density perturbations the initial curvaton VEV is lower, and therefore
$\oosc$ is lower.  As a result the bound is actually stronger by a
factor $\odec^{-1/2}$.

A Gaussian perturbation spectrum, in agreement with observations,
requires $\odec > 10^{-2}$, that is, the curvaton has to account for at
least one percent of the total energy density at the time of
decay~\cite{wands}.

\subsection{Standard scenario}

In the simplest models the curvaton field remains frozen until $H \sim
m_\sigma$, at which point the field starts oscillating in the
potential well.  The curvaton potential is quadratic and $q \sim 1$.
After inflaton decay the background is radiation dominated and $\alpha
= 1/2$.  We will refer to these parameter values as the standard
scenario.  The bound in Eq.~(\ref{bound}) then becomes
\be
H_* \gtrsim A \mpl \lambda^{1/2}.
\label{H_standard_1}
\ee
Here $\Gamma_\sigma \sim \lambda^2 m_\sigma$ and $ \lambda \geq
\max[\lambda_{\rm bbn},\, \lambda_{\rm grav}], $ with 
\bea
\lambda_{\rm bbn} &\sim& \sqrt{H_{\rm bbn}/m_\sigma},  
\label{lambdaBBN}\\
\lambda_{\rm grav} &\sim& m_\sigma/\mpl,
\label{lambdagrav}
\eea
the coupling corresponding to decay just before BBN, and a
gravitational strength coupling respectively.  Plugging in the minimum
value for the coupling gives
\be
H_* > \max \left[
10^7 \GeV \( \frac{H_*}{m_\sigma} \)^{1/5}  
,\,
5 \times 10^{11} \GeV \( \frac{m_\sigma}{H_*} \)
\right]
\label{bnd_ss}
\ee
The first bound in Eq.~(\ref{bnd_ss}) is stronger for $\lambda_{\rm
  bbn} > \lambda_{\rm grav}$ or $m_\sigma < m_{\rm cr} (H_{\rm bbn})
\sim 10^4 \GeV$, whereas in the opposite limit the second bound
applies. Here $ m_{\rm cr}(\Gamma_\sigma)$ is the mass for which the
decay rate becomes of gravitational strength:
\be
m_{\rm cr}(\Gamma_\sigma) = (\Gamma_\sigma \mpl^2)^{1/3}.
\label{mcr}
\ee
The bound is minimized by minimizing the coupling.  For $\lambda =
\lambda_{\rm bbn}$ this means increasing the mass to its largest
possible value; for $\lambda = \lambda_{\rm grav}$ it means decreasing
the curvaton mass as much as possible.  Hence, the bound is weakest in
the limit $m_\sigma \to m_{\rm cr}(H_{\rm bbn})$ for which
$\Gamma_{\rm grav} \to \Gamma_{\rm bbn}$.  The absolute lower bound
then is
\be
H_* > 8 \times 10^7 \GeV.
\ee
This result is clearly incompatible with low scale inflation with
$H_* \sim \TeV$.  

The lower bound is obtained in the limit $\Omega_{\rm dec} \to 1$,
when Eq.~(\ref{rdecay}) is saturated.  It is assumed that inflaton
decay happens before the onset of curvaton oscillations, $\Gamma_I
\gtrsim H_{\rm osc}$, with $I$ denoting the inflaton field.  Otherwise
the lower bound is increased by a factor $(m_\sigma/\Gamma_I)^{1/4}$.

\section{Evading the bound}
\label{s_evading}

To evade, or at least, to weaken the above bound on the Hubble
constant during inflation, one can try one or several of the following
possibilities:
\begin{itemize}
\item
$q > 1$
\item
$\alpha > 1/2$  
\item
$H_{\rm osc}/m_\sigma < 1$
\end{itemize}

A damping factor $q>1$ is possible with a potential that grows faster
than quadratic at large VEV~\cite{damping}. Nevertheless, the bound
will be stronger as the combination $\sigma_* \propto \Omega_\sigma q$
always decreases faster than for a quadratic potential.  Another
possibility is that the potential increases slower than quadratic.
However, small sub-horizon scale fluctuations will grow at a much
faster rate than the super-horizon scale fluctuations observed by CMB
experiments.  The condensate will fragment or decay into $Q$-balls
when $H \sim m_\sigma$~\cite{kusenko}.  $Q$-balls behave as
non-relativistic matter, and the curvaton bound is not much different
from the standard case.

It is possible to have $\alpha>1/2$ if the background is dominated by
a fluid that red shifts faster than radiation and $\omega_{\rm tot} >
1/3$.  If the inflaton energy density is dominated by kinetic energy
the inflaton behaves as a stiff fluid and $\omega_{\rm tot} =1$, giving
an explicit realization.

If the effective curvaton mass increases after the end of inflation it
is possible to have $H_{\rm osc}/m_\sigma < 1$.  Such a mass change
could be triggered by a phase transition --- a Higgs mechanism comes
to mind.  Thermal masses are no good for this purpose, as early
thermal evaporation should be avoided.

\subsection{Stiff inflaton fluid}

The bound in Eq.~(\ref{bound}) is weakened by increasing the exponent
$\alpha > 1/2$.  The maximum value, $\alpha=1$, is obtained for a
stiff fluid with $\omega = 1$.  This is exactly what happens if
inflation is followed by a period of kination, which occurs for
example in quintessential inflation~\cite{peebles,lowquin} and
disformal inflation~\cite{disformal}.  The bound for a background
dominated by a stiff fluid is

\be
H_* > \mpl A \lambda.
\ee
Plugging in the minimal coupling gives 
\be
H_* > \max \left [
10^2 \GeV \( \frac{H_*}{m_\sigma} \)^{1/3}  
,\,
5 \times 10^{-4} m_\sigma
\right ].
\ee
\label{stiff}
For $m_\sigma < m_{\rm cr}(H_{\rm bbn}) \sim 10^4 \GeV$ (see
Eq.~(\ref{mcr})) the first bound is stronger, whereas in the opposite
limit the second expression prevails. A scale invariant spectrum is
obtained for $H_* > 10 m_\sigma$, which is actually a stronger bound
for gravitational decay. The absolute lower bound on the scale of
inflation then is
\be
H_* > 10^2 \GeV.
\ee
This is compatible with low scale inflation, though not very
comfortably. 

In Ref.~\cite{tracking} the bound on the Hubble scale for
gravitational decay --- the second bound on the right hand side of
Eq.~(\ref{stiff}) --- was derived.  However, Ref.~\cite{tracking} then
proceeds by taking $m \to m_{\rm cr}(H_{\rm bbn}) \sim \TeV$ to obtain
the lower bound $H_* > \GeV$, forgetting that $m \gg H_*$ is
incompattible with a scale invariant perturbation spectrum.

It is assumed that inflaton decay happens after the curvaton comes to
dominate the energy density.  Otherwise, if the inflaton decays into
radiation, $\alpha \to 1/2$ as in the standard scenario and the bound
gets stronger by a factor $(\Gamma_I/\Gamma_\sigma)^{1/4}$.  For
$\TeV$ scale inflaton masses and gravitational strength couplings
inflaton decay happens at $\Gamma_I \sim 10^{-4} H_{\rm bbn}$.  This
poses no problem as long as the inflaton energy density is small
enough at the time of decay, so that the accompanying entropy
production is small; in this scenario the universe is reheated by
curvaton decay.  Low scale inflation with $H_* \sim \TeV$ requires
$\Gamma_\sigma /H_{\rm bbn} \lesssim 10^2$.  This constitutes tuning.

\subsection{Heavy curvaton}

In the heavy curvaton scenario the curvaton mass increases
significantly during a phase transition, at $H = H_{\rm pt}$, from
$(m_\sigma)_* \to m_\sigma \gg (m_\sigma)_*$ with $(m_\sigma)_*$ the
curvaton mass during inflation.  If the curvaton mass after the phase
transition is smaller than the Hubble constant $m_\sigma < H_{\rm
  PT}$, oscillations set in when $H_{\rm osc} \sim m_\sigma$ and the
bound of the standard scenario applies, with the sole difference that
now the mass can be larger than the Hubble scale during inflation (but
$(m_\sigma)_* < H_*$ to assure a scale invariant spectrum).

The more interesting case occurs in the opposite limit, $m_\sigma >
H_{\rm pt}$, when oscillations set in at the time of the phase
transition and the ratio $H_{\rm osc}/m_\sigma$ is less than unity.
This may weaken the bound on the inflationary scale Eq.~(\ref{bound})
with respect to the standard scenario.  This possibility was discussed
in Refs.~\cite{lyth,matsuda}.  However, these papers did not take into
account the accompanying increase in the decay rate, and the
constraints this puts on the scenario.

\subsubsection{Mass increase at the end of inflation: 
$H_{\rm pt} \sim H_*$}

We consider first the case that the curvaton mass increases right
after the end of inflation.  The advantage of such a scenario is that
the phase transition triggering the mass increase can be be part of
the inflaton sector, and hence no new dynamics need to be introduced.
This would be realized for example if the curvaton field is coupled to
the waterfall field in hybrid inflation.

As the curvaton mass increases so does the curvaton decay rate.  As
follows from Eq.~(\ref{bound}), a larger mass can lower the bound on
$H_*$, whereas a larger decay rate is counter productive.  The upper
bound on the mass is
\be m_\sigma < \min[m_{\rm en},\, m_{\rm dec}],
\label{mup}
\ee
with
\bea
m_{\rm en} &=& \frac{A H_{\rm pt} \mpl}{H_*} , 
\label {men} \\
m_{\rm dec} &=& \max[\frac{H_{\rm pt}}{\lambda^2},\,
  (\mpl^2 H_{\rm pt})^{1/3}].  
\label{mdec}
\eea
The first bound, $m_\sigma < m_{\rm en}$, is that the energy density
in the curvaton should be less than the total energy in the universe
at the time of the phase transition, $\opt < 1$ with  
\be
\opt \equiv \frac{\rho_\sigma}{\rho_{\rm tot}}\Big |_{H =H_{\rm pt}}
\sim \( \frac{m_\sigma \sigma}{H_{\rm pt} \mpl} \)^2.
\label{rpt}
\ee
If the decay rate increases above $\Gamma_\sigma > H_{\rm pt}$
curvaton decay follows immediately; this sets the upper bound on the
curvaton mass at the time of decay $m_\sigma < m_{\rm dec}$. The first
expression on the right hand side of Eq.~(\ref{mdec}) corresponds to a
coupling constant $\lambda > \lambda_{\rm grav}$, whereas the second
expression is for gravitational strength couplings.

The bound on the inflationary scale is minimized for the largest
possible curvaton mass.  The mass $m_{\rm dec}$ is maximized in the
limit $\lambda \to \lambda_{\rm grav}$.  Direct decay with $\odec \sim
\opt \sim 1$ is only possible if the curvaton dominates the energy
density and $m_{\rm dec} \sim m_{\rm en}$.  With $H_{\rm osc} \sim
\Gamma_\sigma \sim H_*$ and standard scenario values for other
parameters the bound on the Hubble scale during inflation is:
\be 
H_* > A^3 \mpl = 2 \times 10^8 \GeV.
\label{Hpt1}
\ee
This can be found using Eq.~(\ref{bound}) together with the upper
bound on $m_\sigma$ given by Eq.~(\ref{mup}).  Alternatively, one can
solve $m_{\rm dec} \sim m_{\rm en}$ with $\lambda \to \lambda_{\rm
  grav}$.  For smaller values of the Hubble constant $m_{\rm dec} <
m_{\rm en}$ and decay will happen before the curvaton dominates the
energy density.

The above expression is valid for $\Omega_{\rm dec} \sim 1$.  For
smaller values of $\Omega_{\rm dec} < 1$ the mass constraint from
energy conservation is relieved $m_{\rm en} \propto \Omega_{\rm
  dec}^{-1/2}$, while $m_{\rm dec}$ remains unchanged.  As a result,
$m_{\rm dec} < m_{\rm en}$ and decay occurs when the curvaton density
parameter is too small $\Omega_\sigma < \Omega_{\rm dec}$, unless
\be
H_* > \frac{A^3 \mpl}{\Omega_{\rm dec}^{3/2}}.
\ee
The bound on the inflationary scale is minimized in the limit
$\Omega_{\rm dec} \to 1$, and is given by Eq.~(\ref{Hpt1}).  Our
results differ from Lyth~\cite{lyth}, who finds $H_* \gtrsim 10^5
\GeV$.  The reason is that although Lyth considers energy conservation
(he takes $\Omega_{\rm dec} \to 10^{-2}$ to increase $m_{\rm en}$ as
much as possible), he does not take into account the constrains from
direct decay, i.e., the constraint $m < m_{\rm dec}$.

Low scale inflation is excluded.  What is more, this scenario offers
no improvement over the standard scenario, while complicating the
curvaton potential considerably.

\subsubsection{Mass increase after inflation:
  $H_{\rm pt} \ll H_*$}
\label{s_HC2}

From Eq.~(\ref{bound}) it can be seen that the bound on the Hubble
constant can be lowered if 
\be
H_{\rm osc} = \max[H_{\rm pt},\,(m_\sigma)_*]
\label{Hosc}
\ee
is lowered, i.e., if $H_{\rm pt} \ll H_*$.  This was noticed in
Ref.~\cite{matsuda}.  Delaying the onset of oscillations means
lowering $m_{\rm dec}$, see Eq.~(\ref{mdec}), which is counter
productive.  However, the effect on $H_{\rm osc}$ wins, and it is
advantageous to lower $H_{\rm osc}$ as much as possible.  There is no
lower bound on the scale of the phase transition except that it should
take place before curvaton decay, which in turn should occur before
BBN: $H_{\rm pt} > \Gamma_\sigma >H_{\rm bbn}$.

The bound on the Hubble scale during inflation is minimized in the
limit $H_{\rm osc} = H_{\rm pt} \to H_{\rm bbn}$, and direct decay
with $\odec \sim \opt \sim 1$.  After minimizing the onset of
oscillation, the mass $m_{\rm dec}$ should be maximized, which means
$\lambda \to \lambda_{\rm grav}$.  (In this limit $\lambda_{\rm bbn} =
\lambda_{\rm grav}$).  Hence, we find just as in the previous
subsection that the bound is minimized for $\lambda \sim \lambda_{\rm
  grav}$ and $m_{\rm en} \sim m_{\rm dec}$.  This last requirement is
equivalent to demanding the curvaton to decay right after the phase
transition while dominating the energy density. In addition, and in
contrast with the scenario in which the phase transition takes place
promptly at the end of inflation, the bound is further minimized by
taking $H_{\rm osc} \to H_{\rm bbn}$.  The bound on the Hubble
constant then is
\be
H_* \Big|_{\opt \sim 1}>
 \max [\lambda^2 A \mpl, \, A H_{\rm pt}^{2/3} \mpl^{1/3}] >
10^{-14} \GeV.  
\label{Hpt2}
\ee
The equation is saturated for direct decay.  The first bound is
strongest if the coupling $\lambda > \lambda_{\rm grav}$ or $m_\sigma
< m_{\rm cr}(\Gamma_\sigma)$, and vice versa for the second bound.
 
Low scale inflation with $H_* \sim \TeV$ is clearly possible. It
requires $H_{\rm osc} < \GeV$, and thus $(m_\sigma)_*, \, H_{\rm pt} <
\GeV$.  If $(m_\sigma)_* > H_{\rm pt}$, and oscillations set in before
the phase transition, the bound is raised by a factor $((m_\sigma)_*/
H_{\rm pt})$.

If the energy density in the curvaton field after the phase transition
is less than the total energy density, $\opt \ll 1$, direct decay will
lead to an unacceptable level of non-Gaussianity.  A period of
curvaton oscillations is required, long enough for the curvaton to
dominate at decay.  By energy conservation, the curvaton mass is
$m_\sigma = \opt^{1/2} m_{\rm en}$, with $m_{\rm en}$ given in
Eq.~(\ref{men}).  In a radiation dominated universe with $\alpha =
1/2$ the curvaton comes to dominate the energy density at $H_{\rm dom}
= \opt^2 H_{\rm pt}$.  At that moment $\alpha \to 0$, and there is no
further gain in lowering $\Gamma_\sigma< H_{\rm dom}$.  The lower
bound on the Hubble scale during inflation, see Eq.~(\ref{bound}), is
$H_* \propto H_{\rm dom}^{1/4}/m_{\sigma}$ and thus the $\opt$
dependence cancels out.  The bound on the Hubble constant for $\opt <
1$ is the same as for direct decay with $\opt =1$, as long as
domination takes place before decay, that is $\Gamma_\sigma < H_{\rm
  dom} = \opt^2 H_{\rm pt}$.  This latter requirement is actually
stronger and gives
\be
H_* > \max \left[ \frac{A \mpl \lambda^2}{\opt^{3/2}},
\, \frac{A H_{\rm pt}^{2/3} \mpl^{1/3}}{\opt^{1/6}} \right]
\label{Hpt3}
\ee
for a coupling $\lambda \geq \lambda_{\rm bbn},\,\lambda_{\rm grav}$
and for $\lambda = \lambda_{\rm grav}$ respectively.  

Having $\opt <1$ increases the bound for a gravitational decay rate.
The reason is that $\Gamma_{\rm grav} \propto \opt^{3/2} m_{\rm
  en}^3$, and unless the Hubble constant during inflation is higher
than that for direct decay (the value when Eq.~(\ref{Hpt2}) is
saturated) decay happens before domination.  Likewise, for a {\it
  fixed coupling constant} $\lambda > \lambda_{\rm grav}$, the bound
is minimized for direct decay with $\opt \sim 1$. If $\opt < 1$, then
since $\Gamma_\sigma \propto \opt^{1/2} m_{\rm en}$, also here decay
will happen before domination unless $H_*$ is raised above the
saturation value in Eq.~(\ref{Hpt2}).  Another way to interpret the
bound Eq.~(\ref{Hpt3}), however, is that for a {\it fixed inflationary
  scale} $H_*$, the curvaton scenario can equally work for $\opt \sim
1$ and $\lambda^2 < H_*/(A \mpl)$, as for $\opt < 1$ and a smaller
coupling $\lambda^2 < \opt^{3/2} H_*/( A \mpl)$.  In both cases the
curvaton dominates before decay.  The requirement $\max[H_{\rm
  bbn},\,\Gamma_{\rm grav}] < \Gamma_\sigma < H_{\rm dom}$ reads in
terms of the coupling
\be
\max \left[ \(\frac{\opt^{1/2} H_{\rm pt} A}{H_*} \)^2, \,
\frac{H_{\rm bbn} H_*}{\opt^{1/2} A H_{\rm pt} \mpl} \right]
< \lambda^2 < 
\frac{\opt^{3/2} H_*}{\mpl A}.
\label{lambda}
\ee
The minimum bound Eq.~(\ref{Hpt2}) is obtained in the limit $H_{\rm
  BBN} \sim \Gamma_{\rm grav} \sim H_{\rm dom}$, which requires tuning
$\opt =1$ and $\lambda = H_*/(\mpl A)=(H_{\rm pt}/\mpl)^{1/3}$.

Eq.~(\ref{lambda}) gives a lower bound on $\opt$ for the curvaton
scenario to work
\be
\opt > \max\[ \(\frac{H_{\rm bbn}}{H_{\rm pt}} \)^{1/2},\, 
\(\frac{\mpl H_{\rm pt}^2 A^3}{H_*^3} \)^2 \].
\label{betamin}
\ee
For low scale inflation $H_* \sim 10^3 \GeV$, together with
Eq.~(\ref{Hpt3}), this gives $\opt > 10^{-12}$.

\subsection{Additional ingredients}
\label{s_ingredients}

Above we have shown that low scale inflation is compatible with the
curvaton scenario in certain circumstances.  But there are obstacles
every working model has to face, such as the existence of one or more
small masses and late time baryogenesis.  Small couplings are needed
to avoid thermal evaporation.

A small curvaton mass is needed, $(m_\sigma)_* < 0.1 H_*$ for the
stiff inflaton fluid, and $(m_\sigma)_*< 1\GeV$ in the heavy curvaton
scenario.  This is hard to obtain in the presence of
supersymmetry/supergravity, as all scalar fields obtain a soft mass $m
\sim m_{3/2}$ from low energy SUSY breaking, and $m \sim H_*$ from
SUSY breaking by the finite energy density during inflation.  Possible
ways out are tuning (slightly in the case of a stiff inflaton fluid
but considerably for the heavy curvaton); special, non-minimal
K\"ahler potentials which suppress soft masses~\cite{noscale}; global
symmetries (the curvaton as pseudo Nambu Goldstone
boson)~\cite{curv_PNGB}; giving up on supersymmetry.  Soft masses can
be small in gauge mediated SUSY breaking schemes, e.g.  SUSY breaking
at a scale $F \gtrsim (10^5 \GeV)^2$ yields a gravitino mass $m_{3/2}
\gtrsim 10^{-9} \GeV$.  However, one can no longer associate the
inflaton sector with the SUSY breaking sector, or employ moduli fields
to obtain inflation, as this would also imply a very small
inflationary scale $H_* \sim F/\Mpl$.

Baryons and dark matter cannot be created --- the epoch of creation
being the epoch after which their comoving number density remains
constant --- when the curvaton contributes still little to the total
energy density $\Omega_\sigma \ll 1$, as this would give rise to an
unacceptable level of isocurvature perturbations~\cite{wands,gordon}.
Creation by the curvaton field is only consistent with an adiabatic
perturbation spectrum, if the curvaton is close to dominating the
energy density at decay: $\odec > 0.6 \,(0.9)$ for baryons (cold
dark matter)~\footnote{Barring the possibility that baryon production
  before decay leads to an isocurvature perturbation which is exactly
  canceled by an isocurvature perturbation from CDM production after
  decay, or vice versa.}.  There are no constraints if CDM or baryon
production takes place after curvaton decay.  This is a serious
obstacle, especially for small curvaton decay rates and late time
phase transitions $\Gamma_\sigma,\,H_{\rm pt} \to H_{\rm bbn}$, as
there are no known production mechanisms that work at such a late
time.

The stiff inflaton fluid is not affected by thermal effects, as the
reheating temperature is low and all couplings involved are small.
This is not necessarily the case for the heavy curvaton.  Consider a
curvaton coupling $\lambda$ to a standard model (SM) field.  The SM
field is in thermal equilibrium with the radiation bath if $T >
\lambda \sigma$.  Evaporation of the curvaton condensate is avoided if
the scattering rate is sufficiently small: $\Gamma_\sigma \sim \lambda
\alpha_g T < H$ (with $\alpha_g$ a SM structure
constant)~\cite{thermal}, or
\be
H >  \lambda^2 \alpha_g^2 \mpl.  
\label{thermal}
\ee
Note that if this equation is satisfied comfortably, even the case of
direct decay is compromised, as the thermal bath from curvaton decay
will give rise to evaporation before $\Omega_\sigma \sim 1$.

\section{The Higgsed curvaton}
\label{s_higgs}

A straightforward way to implement the heavy curvaton is by
introducing a coupling $h^2 \phi^2 \sigma^2$ in the potential.  The
field $\phi$ has zero VEV during inflation, but acquires a large VEV
during some phase transition at $H = H_{\rm pt} < H_*$.  The effective
mass of the curvaton increases suddenly at this phase transition.

\subsection{The potential}

Consider the following superpotential
\be
W = \kappa S (\phi^2-v^2) + h \sigma (\phi^2 - f^2).
\label{W}
\ee
We take $v = f$ so that the energy density vanishes in vacuum; this
will not influence the results in an essential way.  The $F$-term
potential is
\be
V = (|\kappa|^2 + |h|^2) | \phi^2 - v^2|^2 +4|\phi|^2|\kappa S+h \sigma|^2. 
\label{Vhiggs}
\ee
The potential is of the kind used in hybrid inflation. In analogy we
will call $S$ the flaton (flat direction) field, and $\phi$ the
waterfall field.  The curvaton is denoted by $\sigma$ as before.

$S$ and $\sigma$ are massless in the supersymmetric limit. They have
small masses induced by Planck suppressed operators and/or soft
masses, which we assume are much smaller than the scale of inflation.
During inflation $S$ and $\sigma$ are flat and acquire a large VEV.
The field $\phi$ is heavy due to its coupling with $S$ and remains
trapped at the origin.  At the end of inflation the field $S$ starts
rolling down the potential when $H \sim m_S$.  This is the moment of
the phase transition, since as $S$ approaches zero the effective
$\phi$ mass becomes negative.  The waterfall field rolls down to its
minimum $\phi = v$ and the effective curvaton mass increases due to
its coupling with $\phi$.

The superpotential in Eq.~(\ref{W}) is invariant under a discreet
symmetry under which $Q_S = Q_\sigma = 1$ and $Q_\phi = 0$. This
symmetry is chosen to avoid superpotential terms of the form
\be
W = \lambda_1 S^3 + \lambda_2 \sigma^3 + \lambda_3 \phi^3  
+ \lambda_4 \sigma^2 \phi + \lambda_5 S^2 \phi.
\ee
The first two terms give masses to $S$ and $\sigma$ respectively, and
should be very small.  Remember that $H_{\rm osc} \sim
\max[m_S,\,m_\sigma]$ should be sufficiently small for low scale
inflation to be possible.  The $\lambda_4$ and $\lambda_5$ terms lift
the flatness of the $\sigma$ and $S$ direction by quartic couplings.
Moreover they lead to terms mixing the masses of $\sigma$ and $S$ with
that of the much heavier field $\phi$; these couplings should be
sufficiently small to keep the $\sigma$ and $S$ fields light.
Finally, the $\lambda_3$ term gives rise to a quartic self interaction
for $\phi$, which will be important only if $\lambda_3^2 > h^2 +
\kappa^2$. This coupling does not need to be anomalous small.

Without loss of generality we can take $\phi$ and the couplings
$\kappa,\,\eta$ real.  The only phase left is ${\rm Arg}[S \Sigma]$,
which only shows up in the term $ V \ni 8 h \kappa S \sigma \phi^2 +
{\rm h.c.}$. This term plays no role in the dynamics, it is zero both
before the phase transition when $\phi =0$, and after when $S=0$.  We
will henceforth neglect this term.  Writing the potential in terms of
real fields (denoted by the same symbol), and with the rescalings
$\kappa \to \kappa/\sqrt{2},\,h \to h/\sqrt{2}$, $\eta \to
\eta/\sqrt{2}$ and $v \to v/\sqrt{2}$, the relevant part of the
potential becomes:
\be
V = \frac{\eta^2}{2} (\phi^2 -v^2)^2 + \frac{\kappa}{2} \phi^2 S^2 
+ \frac{h^2}{2} \phi^2 \sigma^2,
\label{V}
\ee
with $ \eta^2 = h^2+\kappa^2$.  The vacuum energy $V_0 \equiv
V(\phi = 0)= v^4 \eta^2/2$.  The effective masses for the curvaton and
waterfall field are
\bea
m_\sigma^2 &=& (m_\sigma)_*^2 + (h v)^2, \\ 
\label{msigma}
m_\phi^2 &=& - (\eta v)^2 + (\kappa S)^2 + (h \sigma)^2,
\label{mphie}
\eea
The contribution of the $h \sigma$-term to the $\phi$-mass is
sub-dominant at all times.  The curvaton coupling to standard model
(SM) particles is not included explicitly in the potential, but is
assumed to be suppressed.

We have chosen for a two-field, hybrid inflation type of potential
rather than a one-field potential to set off the phase transition, as
this offers better perspectives for lowering the bound on $H_*$.
Consider a one-field potential, Eqs.~(\ref{W},~\ref{Vhiggs}) with
$\kappa =0$.  Then the $\phi$ field has to be light, as the phase
transition starts at $H \sim m_\phi$, when $\phi$ starts rolling away
from the origin towards its minimum.  Consequently, the curvaton mass
after the phase transition $m_\sigma = h v =(h/\eta) m_\phi$ remains
small, unless $h/\eta$ is large. But a large $h$ means a large decay
rate $\Gamma ({\sigma \to \phi})$, leading to early curvaton
decay.

Apart from that, a light $\phi$-field has other problems. First of
all, there is no reason for it to have a small VEV after inflation
instead of sitting at its minimum. Secondly, a light field with a
negative mass is hard to obtain in SUSY theories.  Either, the mass
can be driven negative by radiative corrections (but this means large
couplings, and hence, a large decay rate), or the mass term is
dominated by negative soft contributions (implying a non-minimal
K\"ahler potential in gravity mediated SUSY breaking).  Thirdly, the
light $\phi$ field will fluctuate freely during inflation, and $\delta
\rho_\phi/\rho_\phi \sim H_*/\phi_*$ --- which is large and
non-Gaussian for small $\phi_*$.  Its contribution to the total
density perturbations should be negligible small.  There are two ways
in which $\phi$ can imprint its perturbations on the CMB. If
$\rho_\phi$ contributes significantly to the total energy density,
$\phi$ acts itself as a curvaton.  This contribution to the density
perturbations is negligible, less than one percent, if
\be
\ophi \equiv \frac{\rho_\phi}{\rho_{\rm tot}} \Big|_{H= H_{\rm pt}}
< 10^{-2} \( \frac{\phi_*}{\sigma_*} \)
\label{rphi}
\ee
The ratio $\ophi$ is evaluated at the time of the phase transition,
when it has its maximum value (we omit the subscript ``pt'' to avoid
notational clustering).  If the field is initially close to the origin
$\phi_* \sim \delta \phi_* \sim H_*$, this implies $\ophi < 10^{-7}$.
Further, $\phi$ can play the r\^ole of the modulating field in the
inhomogeneous reheating scenario~\cite{dvali}.  Indeed, both the
curvaton mass and decay rate are a function of $\phi$. The density
perturbations produced in this way are $\delta \rho / \rho \sim \delta
\phi / \phi_{\rm dec}$, and are negligible only if the VEV at the time
of curvaton decay is sufficiently large, $\phi_{\rm dec} \gg
\sigma_*$.

All these constraints and problems are avoided simultaneously by
making the $\phi$ field heavy during inflation.  Of course, the
density perturbations of the light field $S$ likewise should be
suppressed.  But this is done much more easily. It requires $\rho_S
\ll  \sigma_*/S_*$; as there is no constraint on the VEV during
inflation one can simply take $S_* \gg \sigma_*$.  The curvaton mass
and decay rate are independent of $S$ at the time of curvaton decay,
and therefore reheating is homogeneous.

\subsection{The bound}
\label{s_bound}

The model parameters have to satisfy several constraints:

\begin{enumerate}
  
\item 
  
  If the potential energy density dominates before the phase
  transition it will drive a period of inflation, hybrid inflation for
  $V_0$ dominance and chaotic inflation for $S$ dominance.  A long
  period of inflation should be avoided as this erases all traces of
  the inflationary period at $H_*$.  This is assured if $\ophi, \,
  \os <1$ or $S_* < \mpl$ and $v^2 \eta <H_{\rm pt} \mpl$.

\item
  
  The mass of the waterfall field should be positive and large during
  inflation, so that it is trapped at the origin and does not
  contribute to the density perturbations: $\kappa S_* > H_*,\,\eta
  v$.  In addition, the density perturbations produced by the flaton
  field $S$ should be sufficiently small, assured if $\os < 10^{-2}
  S_*/\sigma_*$.

\item
  
  There can only be an improvement with respect to standard scenario
  if $m_\sigma = hv \gg (m_\sigma)_*$.  This is only possible if the
  vacuum mass of the waterfall field is much larger than the curvaton
  mass: $m_\phi / m_\sigma \gg 1$.

\item
  
  There is an hierarchy of scales $ H_{\rm bbn} < \Gamma_\sigma <
  H_{\rm pt}$; the curvaton should decay before BBN, but after the
  phase transition.  $\Gamma_\sigma$ is the decay rate into SM matter,
  which should exceed curvaton decay into any of the hidden sector
  fields, such as $\phi$ and $S$. Since $m_\sigma < m_\phi$ after the
  phase transition, curvaton decay into $\phi$ quanta is kinematically
  inaccessible.  However, if the fermionic superpartners of $\phi$ are
  sufficiently light, the curvaton can still decay into them and
  $\Gamma_\sigma > h^2 m_\sigma$ is needed.

\end{enumerate}

\noindent
The parameters $\opt$ and $\ophi$, defined in
Eqs.~(\ref{rpt},~\ref{rphi}), give rise to the equalities
\be
\rho_\sigma/\opt  = \rho_\phi/\ophi = \rho_{\rm tot}
\label{master}
\ee
Here $\rho_\phi = V_0$ is the potential energy density stored in the
waterfall field before the phase transition, $\rho_\sigma = (m_\sigma
\sigma_*)^2$ is the energy density transferred to the curvaton field
during the phase transition, and $\rho_{\rm tot} = (H_{\rm pt}
\mpl)^2$ is the total energy density at the time of the phase
transition.  Further $\opt \leq \ophi$, saturated when the energy
stored in $V_0$ is transferred entirely to the curvaton field.  From
the above equation we can solve
\bea
m_\sigma &=& \frac{\opt^{1/2} H_{\rm pt} \mpl}{\sigma_*} = 
\( \frac{\ophi^{1/2} H_{\rm pt} \mpl}{\eta} \)^{1/2}
\label{mass}\\
\eta &=& \frac{\ophi^{1/2}}{\opt}
\frac{(h \sigma_*)^2}{H_{\rm pt} \mpl}
\label{eta}
\eea
From the first equality in Eq.~(\ref{mass}) it seems that direct decay
with $\opt \sim \ophi \to 1$ is possible for all $H_{\rm pt}$, and
thus the bound on the Hubble constant during inflation can be $H_* >
10^{-14} \GeV$, the bound obtained in section~\ref{s_HC2}.  However,
this would require couplings exceeding unity.  Imposing $h,\, \eta <1$
together with $\eta = \sqrt{h^2 + \kappa^2} > h$~\footnote{If one
  considers the potential in Eq.~(\ref{V}) as an effective potential,
  not coming from a superpotential of the form Eq.~(\ref{W}), then
  there is no relation between the couplings.  One can take the limit
  $\eta < h$ to try to lower the bound.  However, this implies
  $m_\sigma > m_\phi$ and therefore a large decay rate $\Gamma_\sigma
  = h^2 m_\sigma$, which is counter productive.}, the bound for direct
decay with $\opt \sim 1$ is
\be
H_* \Big|_{\opt \sim 1} 
>  \frac{\sqrt{\eta}}{h}  A \sqrt{H_{\rm pt} \mpl}
>
4 \times 10^{-7} \GeV,
\label{bnd_higgs}
\ee
where in the last step we have taken $H_{\rm pt}, \, \Gamma_\sigma \to
H_{\rm bbn}$ and $\sqrt{\eta}/{h} \to 1$.  This bound can be either
found by using Eqs.~(\ref{Hpt2},~\ref{lambda}) with 
\be
\lambda^2 = \frac{H_*}{A \mpl} = \frac {h} {\sqrt{\eta}} 
\( \frac {H_{\rm pt}} {\mpl} \)^{1/2},
\ee
or from the expression for $\eta$ in Eq.~(\ref{eta}) above. As
discussed in section \ref{s_HC2}, the bound for $\opt <1$ is the same
as for direct decay with $\opt =1$, as long as the curvaton density
parameter at the end of the phase transition is large enough and the
decay rate sufficiently small --- see
Eqs.~(\ref{Hpt3},~\ref{lambda},~\ref{betamin}) --- so that curvaton
domination occurs before decay, i.e., $\Gamma_\sigma < \opt^2 H_{\rm
  pt}$.

The bound on the Hubble constant depends on the coupling values.  The
lower bound above is obtained for $h \sim \kappa \sim 1$ and $\lambda
\to \lambda_{\rm bbn} \sim 10^{-11} \gg \lambda_{\rm grav}$.  Then low
scale inflation with $H_* \sim \TeV$ is possible for $({m}_\sigma)_*
,\, H_{\rm pt} < 10^{-6} \GeV$, $m_\phi < 10^6 \GeV$ and $\opt >
10^{-10}$.

It could be argued though, see point (4) above, that the coupling $h
\lesssim \lambda$.  Taking all couplings approximately equal $h \sim
\kappa \sim \lambda \to \lambda_{\rm bbn} \sim 10^{-9}$ gives the
bound $H_* > 10^{-2} \GeV$.  This is still compatible with low scale
inflation, but the amount of tuning is increased.  For example, TeV
scale inflation is only possible for $(m_\sigma)_* ,\, H_{\rm pt} < 5
\times 10^{-15} \GeV$, ${m}_\phi < 10^{-4} \GeV$ and $\opt > 10^{-6}$.
Moreover, these parameter choices violate Eq.~(\ref{thermal});
the curvaton scenario is jeopardized by thermal evaporation.

If the decay rate into SM matter is of gravitational strength curvaton
decay before nucleosynthesis requires $\opt^{1/2} H_{\rm pt} > (H_{\rm
  bbn}/\mpl)^{1/3} \sigma_* > 10^{-8} \GeV$ for low scale inflation
with $H_* \sim 10^3 \GeV$.  From Eq.~(\ref{Hpt3}) we also get an upper
bound on the time of the phase transition $H_{\rm pt} <
10^{-6}/\opt^{1/4} \GeV$.  There is only a small window of $H_{\rm
  pt}$ for which the heavy curvaton scenario can work; in addition
$\opt > 10^{-3}$ is needed.

\subsection{discussion}

Hubble constants much larger than the theoretical, model independent,
lower bound $H_* > 10^{-14} \GeV$ are needed in the Higgsed curvaton
scenario.  The reason is that the parameters are not all independent.
Tuning both $\ophi \to 1$ and $\opt/\ophi \to 1$ is impossible for
coupling constants not exceeding unity.

The model has the same general problems discussed in section
\ref{s_ingredients}: Late time baryogenesis and at least two small
masses.  More generically there are three small masses, but with a
distinct hierarchy $H_* \gg m_\phi \gg (m_\sigma)_*,\, m_S$.  None of
the light fields can be a pseudo Nambu-Goldstone boson (PNGB) as the
potential terms $\kappa^2 S^2 \phi^2$ and $h^2 \sigma^2 \phi^2$ break
the shift symmetry of all fields.

\section{The heavy curvaton as axion}
\label{s_axion}

Axions are natural curvaton candidates; the axion mass is protected
by a (weakly broken) global symmetry, and therefore can be kept light
naturally, even during inflation when supergravity corrections
generically lead to masses of the order of the Hubble constant for all
scalars.  

\subsection{Axion with mass from non-renormalizable operators}
\label{s_axion_NR}

We consider a potential of the form $V = V_{\rm PQ} + V_{\rm NR}$ with
\be
V_{\rm PQ} = \lambda_\Sigma^2 (\Sigma^2 - f^2)^2.
\ee
The VEV of $\Sigma$ breaks the PQ-symmetry spontaneously at a scale
$\langle \Sigma \rangle = f$.  The curvaton $\sigma$ is identified
with the canonically normalized axion field, defined as $\Sigma =
(f/\sqrt{2}) \exp(i \sigma/f)$.  If the non-renormalizable potential
breaks the PQ symmetry explicitly it gives a mass to the axion
$m_\sigma \propto f^a$, with $a$ a model dependent constant.
Therefore, one way to implement the heavy curvaton is to have an
increase in the VEV of $\Sigma$ at $H_{\rm pt}$~\cite{lyth}. This can
be obtained by a potential term $h \Sigma^2 \phi^2$, with $h < 0$.
Although negative couplings are possible in SUSY theories, they
require very elaborate constructions~\cite{stewart}.  We will
therefore pursue another option.
 
Consider the non-renormalizable potential with operators of dimension
$D_1 =2m_1 + n_1$ and $D_2 = 2l_2 + 2m_2 + n_2$ respectively
\be
V_{\rm NR} = 
\lambda_1^2 \( 
\frac{|\Sigma|^{2m_1} \Sigma^{n_1}}{\mpl^{2m_1+n_1-4}} + {\rm c.c.} \)
+ \lambda_2^2 \( 
\frac{|\phi|^{2l_2} |\Sigma|^{2m_2} \Sigma^{n_2} }{\mpl^{2l_2 +2m_2 +n_2 -4}} 
+ {\rm c.c.} \).
\label{VNR}
\ee
This potential gives two contributions to the curvaton mass
$m_\sigma^2 = m_1^2 + m_2^2$ with
\bea
m_1^2 &=& \lambda_1^2 \( \frac{f}{\mpl} \)^{D_1-2} \mpl^2, \\
m_2^2 &=& \lambda_2^2 \( \frac{f^{D_2-2l_2-2} \phi^{2l_2}}{\mpl^{D_2-2}} 
\) \mpl^2 .
\eea
During inflation $\phi$ has a small VEV and $(m_\sigma)_* = m_1$.
After inflation there is a phase transition in which the VEV of $\phi$
grows to large values and $m_\sigma \approx m_2$.  The bound on the
inflationary scale can be weakened only if $m_2 > m_1$ or
\be
\( \frac{\phi}{f} \)^{2l_2} > \(\frac{\lambda_1}{\lambda_2} \)^{2} 
\(\frac{f}{\mpl}\)^{D_1 - D_2}.
\ee
For operators of the same dimension $D_1 =D_2$ and couplings
$\lambda_1,\, \lambda_2 \sim {\mathcal O}(1)$ this is just the
requirement that $\phi > f$.

In the following we will assume that the initial curvaton VEV is of
the order of its maximal value $\sigma_* \sim f$; this assumption
minimizes the bound on $H_*$.  Curvaton decay into SM matter is
mediated by non-renormalizable operators, so that the PQ symmetry is
preserved at the normalizable level.  The decay rate then is of
gravitational strength and the bound is given by Eq.~(\ref{Hpt3})
\be 
H_* > \frac{ A \mpl^{1/3} H_{\rm pt}^{2/3}}{\opt^{1/6}}.
\label{maxhpt}
\ee
Direct decay corresponds to the limit $\opt \to 1$ and $f \to
\sigma_*$.  The value of $H_{\rm pt}$ cannot be lowered/raised
arbitrarily in this model.  First of all $H_{\rm pt}> m_1$, otherwise
the bound is increased by a factor $(m_1/H_{\rm pt})$. Further $m_2
\propto H_{\rm pt}$, and thus so is the decay rate; $H_{\rm pt}$
should be large enough for decay to take place before BBN.  Lastly
${m}_\phi \geq H_{\rm pt}$. This together constrains
\be
\max \left [\lambda_1\( \frac{f}{\mpl} \)^{\frac{D_1-2}{2}} \mpl,\, 
 \frac{f}{\sqrt{\opt}} \(\frac{H_{\rm bbn}}{\mpl} \)^{1/3} \right] 
<   H_{\rm pt} < 
m_\phi.
\label{minhpt}
\ee
If $D_1 \geq 9$ the limit $H_{\rm pt} \to H_{\rm bbn}$ is possible and
direct decay with $\opt = \ophi \to 1$ gives the theoretical lower
bound $H_* > 10^{-14} \GeV$ of section \ref{s_HC2}.  However, we still
have to add the explicit form of the potential $V_\phi$ that is to
trigger the phase transition.  And as for the Higgsed curvaton, we
expect that this will give additional constraints on the system.  We
will discuss two cases below focusing on the possibility of low scale
inflation with $H_* \sim 10^3\GeV$.  Note that independently of the
potential, the lower bound on $\opt$ from combining
Eqs.~(\ref{maxhpt},~\ref{minhpt}) is
\be
\opt > 10^{-12} \( \frac{H_*}{10^3 {\rm GeV}} \)^{-2/3}.
\label{rptbound}
\ee

\subsubsection{$V_\phi$ quadratic}
\label{s_quadratic}

To achieve the phase transition we add a quadratic potential for $\phi$ 
\be 
V_\phi = - m_\phi^2 \phi^2.  
\ee
The phase transition takes place at $H_{\rm pt} \sim m_\phi$ when the
field starts rolling towards large VEVs until the potential is lifted
by the terms in $V_{\rm NR}$.  The $\phi$-field is necessarily light;
its contribution to the density perturbations is negligible small if
Eq.~(\ref{rphi}) is satisfied.

Assume that at large $\phi$ the dominant term in $V_{\rm NR}$ has
$l_2>2$, so that the potential increases faster than quadratically and
there is no runaway behavior, as well as $n_2 \geq 2$ so that it
generates a mass term for the curvaton.  This is the most advantageous
situation for the heavy curvaton, since then all of the potential
energy is transferred to the curvaton field and $\opt = \ophi$, with
$\opt,\, \ophi$ defined in Eqs.~(\ref{rpt},\,\ref{rphi}).  This gives
the relation $m_2 f \sim m_\phi \phi = \sqrt{\opt} H_{\rm pt} \mpl$,
which can be solved for $\opt$:
\be
\opt  \sim \( \frac{H_{\rm pt}}{\lambda_2 \mpl} \)^{\ts \frac{2}{l_2-1}} 
\(\frac{\mpl}{f} \)^{\ts \frac{D_2-2l_2}{l_2-1}}.
\ee
Eqs.~(\ref{maxhpt},~\ref{minhpt}) put an upper/lower bound on $H_{\rm
  pt}$, and therefore likewise on $\opt$. Taking $\lambda_2 \sim
{\mathcal O}(1)$ we get
\be
\( \frac{\mpl}{f} \)^{\ts \frac{D_2-2l_2-2}{l_2}} 
\( \frac{H_{\rm bbn}}{\mpl} \)^{\ts \frac{2}{3l_2}} 
<\opt<
\( \frac{\mpl}{f} \)^{\ts \frac{2(D_2-2l_2-3)}{2l_2-3}} .
\ee
In Table 1 below the values of $\opt$ are tabulated for various
values of $D_2$ and $l_2$ for the case of low scale inflation with
$H_* \sim \TeV$ and $f \sim \sigma_* = H_*/A$.
$$
\stackrel{
\begin{array}[t]{|c||c|c|c|}\hline
(D_2 ,\, l_2) & (\opt)_{\rm min} & (\opt)_{\rm max}  & \delta D\\
\hline \hline
(6 ,\, 2) &  - & - & -\\ \hline
(7  ,\, 2) & 6 \times 10^{-9} & 1 & 2 \\ \hline
(8  ,\, 2) & 6 \times 10^{-3} & 1  & 18\\ \hline
(8 ,\,  3) & 9 \times 10^{-9} &  3 \times 10^{-10} &2  \\ \hline
(9 ,\,  2) & - &  - &- \\ \hline
(9 ,\,  3) & 3 \times 10^{-6} &  1 & 6  \\ \hline
(10 ,\,  2) & -  &  -  & - \\ \hline
(10 ,\,  3) & 3 \times 10^{-2} &  1 & 40  \\ \hline
(10 ,\,  4) & 7 \times 10^{-8} &  1 \times 10^{-5} &4 \\ \hline 
\end{array}}{\stackrel{}{
\mbox{Table 1} }} $$

The heavy curvaton cannot work for $(D_2,\,l_2) = (6,2), \, (9,2),
\,(10,2)$ as these operators cannot give $\opt > 10^{-12}$, see
Eq.~(\ref{rptbound}).  A low value $\opt <10^{-7}$, in order to
suppress the $\phi$ contribution to the density perturbations, is only
possible for $(D_2,\,l_2) = (7,2),\,(8,3)$ and $(10,4)$.  The mass
$(m_\sigma)_* = m_1 < (H_{\rm pt})_{\rm min} $ for all $D_1 \geq 7$,
in agreement with Eq.~(\ref{minhpt}).

The dominant contribution to the curvaton mass is provided by a
dimension $D_2$ operator with $l_2 \geq 2$.  In addition there might
be operators in $V_{\rm NR}$ which are pure monomials in $\phi$ (terms
with $D_2 = 2l_2$ in the language of Eq.~(\ref{VNR})); call the
dimension of the lowest order such operator $D_\phi$. If this latter
term dominates the potential at large $\phi$, then only part of the
potential energy stored in $\phi$ is transferred to curvaton energy
during the phase transition and $\opt < \ophi$.  For $D_\phi = D_2$
the $\phi$ monomial always dominates.  In this case $\opt = \ophi^2
V_{D_2} / V_{D_\phi} \ll 10^{-12}$ for all values of $(D_2,\,l_2)$;
the heavy curvaton scenario does not work, as follows from
Eq.~(\ref{rptbound}). The $\phi$-monomial is sub-dominant, and the
curvaton scenario works as before, if $D_\phi = D_2 + \delta D$ with
$\delta D \geq 2$.  The values of $\delta D$ for the different
combinations of $(D_2,\,l_2)$ are given in Table 1 above.

The favored operators, those with a low value of $\delta D$, are the
same operators that give a low value of $\opt$.  Not surprisingly,
these are the operators with the largest value of $l_2$, maximizing
the mass increase $(m_2/m_1)$.

\subsubsection{$V_\phi$ quartic}

In this subsection we take the potential responsible for the phase
transition to be quartic, of the form
\be
V_\phi = -m_\phi^2 \phi^2 + \eta^2 \phi^4.
\ee
If this is all, and $H_{\rm pt} \sim m_\phi$ as in the previous
subsection, there is no gain in introducing the quartic term as it can
only be harmful and lead to $\opt < \ophi$.  So we assume that the
potential is embedded in a hybrid inflation type of potential of the
form Eq.~(\ref{V}).  If this potential originates from the
superpotential in Eq.~(\ref{W}), then the quartic coupling is
restricted: $\eta > H_*/ S_* > 10^{-15}$ for $H_* \sim \TeV$, see
point (2) in section \ref{s_bound}.  In this set up it is possible to
have $m_\phi > H_{\rm pt}$, which gives more freedom in model
building. Further, the $\phi$-field is now heavy during inflation and
does not fluctuate; consequently the upper bound on $\opt$ of
Eq.~(\ref{rphi}) does not apply.

For large coupling $\eta >10^{-6 (D_2-4)}$, the potential is lifted by
the quartic term at large $\phi$ and not by the operators in $V_{\rm
  NR}$, and the VEV is $\phi = m_\phi/\eta$.  We can solve for $\opt =
{V_{\rm NR}}/(H_{\rm pt} \mpl)^2$, giving
\be
\opt = \lambda_2^2 \( \frac{H_{\rm pt}}{\mpl} \)^{2(l_2-1)}
\( \frac{f}{\mpl} \)^{D_2-2l_2} 
\( \frac{m_\phi}{H_{\rm pt} \eta}\)^{2l_2}.
\ee
Using the upper/lower bound on the scale of the phase transition, see
Eqs.~(\ref{maxhpt},~\ref{minhpt}), we again get an upper/lower bound
on $\opt$:
\be
\( \frac{f}{\mpl} \)^{\ts \frac{D_2-2}{l_2}} 
\( \frac{H_{\rm bbn}}{\mpl} \)^{\ts \frac{2(l_2-1)}{3l_2}}
 \( \frac{m_\phi}{H_{\rm pt} \eta}\)^2
<\opt<
\( \frac{f}{\mpl} \)^{\ts \frac{2(D_2+l_2-3)}{3-l_2}}  
\( \frac{m_\phi}{H_{\rm pt} \eta}\)^{\ts \frac{4l_2}{3-l_2}}
\label{l2is4}
\ee
The lower bound is generically small $(\opt)_{\rm min} \ll 1$, and can
easily be satisfied.  The whole issue is whether the upper bound can
be large enough, $\opt > 10^{-12}$ for TeV scale inflation.  The upper
bound is maximized in the limit of small $H_{\rm pt}$ (or equivalently
$\Gamma_\sigma \to H_{\rm bbn}$) and $\eta$.  The $\phi$ mass is
related to the scale of the phase transition by the relation
$m_\phi^2/\eta = \ophi^{1/2} H_{\rm pt} \mpl$.

In Table 2 below we have indicated the values of $(D_2,\,l_2)$ for
which $\opt \sim 1$ is possible. Large $\opt$ is not possible for
$l_2=1$, independent of the value of $D_2$, so we left that option
out. 

$$
\stackrel{
\begin{array}[t]{|c||c|c|}\hline
(D_2 ,\, l_2) & \opt \sim 1 & \beta_0  \\
\hline \hline
(6 ,\, 2)   & \surd & 10^{-8}-10^{-10}\\ \hline
(7  ,\, 2)  & \surd & 10^{5}-10^{-6} \\ \hline
(8  ,\, 2)  & - & - \\ \hline
(8 ,\,  3)  & \surd &  10^{-9}  \\ \hline
(9 ,\,  2)  & - &  -  \\ \hline
(9 ,\,  3)  & \surd &  10^{-5}  \\ \hline
(10 ,\,  2) & - &  -  \\ \hline
(10 ,\,  3) & \surd &  10^{-8}  \\ \hline
(10 ,\,  4) & \surd &  < 10^{-8} \\ \hline
\end{array}}{\stackrel{}{
\mbox{Table 2} }} 
$$

\noindent
Also listed is the boundary value of $\beta = (m_\phi/H_{\rm pt} \eta)
/ (m_\phi/H_{\rm pt} \eta)_{\rm max}$ for which the curvaton scenario
can work. Here
\be
\( \frac{m_\phi}{H_{\rm pt} \eta} \)_{\rm max} =
\left \{ 
\begin{array} {lll}
& 1 \times 10^{19} 
\( \frac{\opt \ophi}{1} \)^{1/4} 
\( \frac{10^{-12}} {\eta}\)^{1/2} 
\( \frac{H_{\rm bbn}}{\Gamma_\sigma} \)^{1/6} ,
& \qquad D_2 =6, \\
& 4 \times 10^{20} 
\( \frac{\opt \ophi}{1} \)^{1/4} 
\( \frac{10^{-15}} {\eta}\)^{1/2} 
\( \frac{H_{\rm bbn}}{\Gamma_\sigma} \)^{1/6} ,
& \qquad D_2  > 6 .
\end{array}
\right.
\ee
For $D_2 =6$ the quartic coupling has to be $\eta >10^{-12}$ for the
quartic term to dominate over $V_{\rm NR}$, whereas for $D_2 > 6$ the
dominant constraint is $\eta > 10^{-15}$ from the requirement that
$\phi$ is heavy during inflation.

For $l_2 =2,\,3$ the curvaton scenario can work for $\beta_0 < \beta <
1$, with $\beta_0$ the value listed in Table 2.  The smaller $\beta_0$
the larger the parameter space, as there is more room for $\eta$,
$\opt$, $\ophi$ and/or $\Gamma_\sigma$ to vary.  For $l_2=2$ the exact
value of $\beta_0$ depends on $\opt$, and there is a $\beta_0$-range:
the upper bound corresponds to $\opt\sim 1$, whereas the lower bound
is for $\opt= \beta$.  Note that the upper bound on the curvaton
density parameter after the phase transition is $\opt > \beta_0$.

For $l_2 \geq 4$ the upper bound on $\opt$ is maximized for large
$\eta$ and $\Gamma_\sigma$, as follows from Eq.~(\ref{l2is4}).  The
value of $\beta_0$ in this case is an {\it upper} bound. For example,
for $(D_2,\,l_2) = (10,\,4)$ one needs $\beta < 10^{-8}$.  In this
case couplings $\eta \sim 1$ are possible.  The lower bound on $\opt$
of Eq.~(\ref{rptbound}) can be reached.

Non-renormalizable terms with $D_\phi = 2l_2$ are negligible small, and
therefore do not alter the above picture, if $\eta >
\lambda_\phi(\phi/\mpl)^{(D_\phi-4)/2}$ or
\be 
\eta > \lambda_\phi 
\( \frac{\ophi^{1/2} H_{\rm pt}}{\mpl} \)^{\ts \frac{D_\phi -4}{D_\phi}}
>  ( 10^{-26} )^{\ts \frac{D_\phi -4}{D_\phi}},
\ee
where in the last step we have taken $\lambda_\phi \sim 1$, $\opt \to
\ophi$ and $H_{\rm pt} \to 10^{-8} \opt^{-1/2}$ (which is the lower
bound of Eq.~(\ref{minhpt})).  The non-renormalizable potential is
negligible for $\eta > 10^{-9},\, 10^{-13}, 10^{-16}$ for $D_\phi =
6,\,8,\,10$.  Even for $D_\phi = D_2$, only part of parameter space is
excluded.  This is in sharp contrast with the quadratic model
discussed in the previous subsection, where $D_\phi - D_2 \geq 2$ is
needed.

\subsection{discussion}

The main difference between the Higgsed curvaton and the axionic
curvaton is that the latter obtains its mass from non-renormalizable
operators instead of through a direct coupling.  The advantage is that
the curvaton can be kept light naturally.  Moreover, since the
$\phi\sigma$-coupling is of gravitational strength, there is no strong
constraint from curvaton decay into $\phi$ or its superpartners.  But
the use of non-renormalizable operators also has it draw backs. There
is less freedom in tuning couplings.  In particular, the curvaton
decay rate is of gravitational strength, which constrains the model
considerably. Further, it is harder to suppress or forbid unwanted
operators, as discrete symmetries might be broken by gravity ---
indeed, the non-renormalizable potential violates the PQ symmetry
explicitly.  This is especially a problem for the quadratic potential
of section 5.2.1.

\section{Conclusions}

We revisited the bound on the Hubble scale during inflation in the
context of the curvaton scenario.  The bound can be weakened in
non-standard settings.  One specific example is the heavy curvaton
scenario.  In this scenario the curvaton mass increases significantly
after the end of inflation, (possibly) triggered by a phase transition.
We reanalyzed the bound in this set up, exhibiting explicitly the
dependence on various parameters.  We take into account the upper
bound on the curvaton mass coming from both energy conservation and
direct decay.  A lower bound on the Hubble constant of $H_* > 10^{-14}
\GeV$ is obtained, in the limit that the mass increase happens just
before nucleosynthesis.  TeV scale inflation requires $H_{\rm pt} <
1\GeV$.

We discussed two implementations of the heavy curvaton in detail.  The
Higgsed curvaton receives a mass through a Higgs mechanism, whereas
the axionic curvaton receives its mass from non-renormalizable
operators.  Although in theory a Hubble constant $H_* \to
10^{-14}\GeV$ is possible, it turns out that even obtaining TeV scale
inflation is hard to achieve.  The reason is that in these explicit
models the various parameters cannot all be varied independently.

\section*{Acknowledgments}
The author is supported by the European Union under the RTN contract
HPRN-CT-2000-00152 Supersymmetry in the Early Universe.

%%%%%%%%%%%%%%%%%%%%%%%%%%%%%%%%%%%%%%%%%%%%%%%%%%%%%%%%%%%%%%%%%%%%%%%%%%
%%%%%%%%%%%%%%%%%%%%%%%%%%%  bibliography  %%%%%%%%%%%%%%%%%%%%%%%%%%%%%%%
%%%%%%%%%%%%%%%%%%%%%%%%%%%%%%%%%%%%%%%%%%%%%%%%%%%%%%%%%%%%%%%%%%%%%%%%%%

\end{document}